\documentclass{ws-ijmpd}
\usepackage[super,compress]{cite}
\usepackage{amsmath}
\usepackage{adjustbox}
\usepackage{multirow}
\usepackage{graphicx}
\usepackage{booktabs}
\usepackage{hyperref}
\usepackage{float}
\usepackage[utf8]{inputenc}

\usepackage{flushend}
\usepackage{newtxtext,newtxmath}
\usepackage{soul}
\usepackage{hyperref}
\usepackage{float}
\usepackage{ulem} 
\usepackage{graphicx}
\usepackage[T1]{fontenc}
\usepackage{ae,aecompl}
\usepackage{amsmath}
\usepackage{amssymb}
\usepackage{scalerel}
\usepackage{tikz}
\usetikzlibrary{svg.path}
\definecolor{orcidlogocol}{HTML}{A6CE39}

\begin{document}

\title{Imprints of the post recombination dissipation of helical magnetic field on the Cosmic Microwave Background Radiation}

\author{SANDHYA JAGANNATHAN}
\address{Department of Physics and Astrophysics, University Of Delhi \\
Delhi  110007,
India\\
sjagannathan@physics.du.ac.in}

\author{RAMKISHOR SHARMA}
\address{IUCAA, Post Bag 4, Pune University Campus, \\ Ganeshkhind, Pune 411007, India \\
ramkishor@iucaa.in}

\author{T.R. SESHADRI}
\address{Department of Physics and Astrophysics, University Of Delhi \\
Delhi  110007,
India\\
trs@physics.du.ac.in}

\maketitle

\begin{history}
\received{Day Month Year}
\revised{Day Month Year}
\end{history}

\begin{abstract}
Astrophysical magnetic fields decay primarily via two processes namely, ambipolar diffusion and turbulence. Constraints on the strength and the spectral index of non-helical magnetic fields have been derived earlier in the literature through the effect of the above mentioned processes on the Cosmic Microwave Background (CMB) radiation. A helical component of the magnetic field is also produced in various models of magnetogenesis, which can explain larger coherence length magnetic field. In this study, we focus on studying the effects of post recombination decay of maximally helical magnetic fields through ambipolar diffusion and decaying magnetic turbulence and the impact of this decay on CMB. We find that helical magnetic fields lead to changes in the evolution of baryon temperature and ionization fraction which in turn lead to modifications in the CMB temperature and polarization anisotropy. These modifications are different from those arising due to non-helical magnetic fields with the changes dependent on the strength and the spectral index of the magnetic field power spectra.  
\end{abstract}

\keywords{Helical magnetic fields, Ambipolar diffusion, Decaying magnetic turbulence, Cosmic Microwave Background radiation}

\ccode{PACS numbers:}

\section{Introduction}

Magnetic fields are observed at various length scales in the universe \cite{r-beck,beck2013,malik2017,bernet2008,clarke,govoni2004,vogt2005,widrow,neronov,taylor2011,go2019}. However, the origin of these fields is still an open problem and there is no clear consensus on whether the large scale magnetic fields observed in galaxies and galaxy clusters are of primordial or of astrophysical origin. Magnetic fields of the order of $10^{-15}$G found using $\gamma$-ray observations of TeV-Blazars in the intergalactic medium~(IGM) indicates that at least a certain fraction of the magnetic field originates from the former~\cite{neronov,taylor2011,kandu2016,tav2011,tak2013,meyer2016,long2015}. Recent observation of magnetic fields located in a filament between two merging galaxy clusters also provides a particularly strong evidence of a primordial origin~\cite{go2019}. There are various mechanisms of magnetogenesis in the early universe and these include, inflation, phase transitions among others~\cite{turner-widrow,ratra,martin-yokoyama,agullo2013,atmjeet2014,Campanelli2015,vachaspati,Sigl:1996dm,kisslinger,qcd,arun2015,zhang2019,ellis2019}. Inflation provides a natural mechanism to generate large scale magnetic fields. However, there are certain problems associated with the same namely, strong coupling and backreaction. Various models have been suggested to circumvent these two issues and generate magnetic fields that satisfies the current observational bonds~\cite{rajeev2013,1475-7516-2015-03-040,sharma2017}. Some of these models, however, require low scales of reheating~\cite{sharma2017,sharmahelical,kobayashi2014}. The magnetic fields evolving from causal processes such as phase transitions do not suffer from the problems that affect inflationary magnetogenesis but they generate magnetic fields that have correlation length much smaller than the Hubble horizon at that epoch. Hence, these fields when evolved till today are correlated on galactic scales or less. However, if a parity violating magnetogenesis mechanism is considered, then magnetic fields with non-zero helicity are produced, which have larger coherence lengths at present along with a slower decay rate compared to non-helical magnetic fields~\cite{kandu2016,jedamzik,rajeev2010,caprini2014,fujita2019,atmjeet2015,chris2005,sav2013,kahn2017,brand2018,brand2015}. In particular, conservation of magnetic helicity sets constraints on the decay time-scale of helical magnetic fields leading to an inverse cascade of energy. The latter implies that the magnetic field power at smaller scales is transferred to larger scales contrary to the process of direct cascade as observed in the case of non-helical magnetic fields. Magnetic helicity has also been invoked in inflationary magnetogenesis models to explain higher magnetic field strength and correlation length for the same value of reheating temperature~\cite{sharmahelical}. \par
Magnetic fields impact various processes in the universe such as structure formation~\cite{wass1978,kim1994,ss2005}, temperature and polarization anisotropies of the Cosmic Microwave Background radiation (CMB) and bounds on such fields have been placed accordingly by studying the clustering of galaxies and CMB observations~\cite{bar1997,durr1999,sesh2001,sub2003,sesh2009,caprini2009, shaw2010, tri2010,tri2012,tri2014,shi2014,pogo2018,pao2019}. The effect of these fields has also been observed in the observations of CMB brightness temperature fluctuations produced by the neutral hydrogen 21cm line~\cite{sch2009,kunze2018,min2019}. Further, the presence of helicity leads to parity odd cross correlations between the CMB temperature and B- polarization anisotropies and E-and B-polarization anisotropies~\cite{pogo2001,caprini2004,kahn2005,ball2015,planck2015}. These parity odd signals provide a clear signature of stochastic helical magnetic fields. The CMB angular power spectra and Planck likelihoods constrain the maximally helical magnetic field today, $B_{0}$, smoothed on a $1$Mpc length scale to be $<5.6$nG by incorporating the contribution of magnetic fields in cosmological perturbations~\cite{planck2015,kandu2016}. Whereas the constraint on non-helical magnetic fields is $<4.4$nG using Planck 2015 data~~\cite{planck2015,kandu2016}. Constraints on non-helical magnetic field strength have been improved  to $<3.4$nG using Planck 2018 data and by further adding BICEP/KECK data \cite{bicep} and South Pole Telescope (SPT) polarization data \cite{spt}, it has been improved  to $<3.5$nG and $<2.9$nG, respectively~\cite{pao2019}. These constraints are further expected to modify when the effects of magnetic field on heating and ionization of the IGM are considered~\cite{ss2005,chluba2015,kunze2014,kunze2015}.    \par
In the standard model of post-recombination universe, the temperature of the IGM follows the CMB temperature up to a redshift of $z \approx 100$ after which it thermally decouples from CMB and evolves as $a^{-2}$ at low redshifts~\cite{kandu2016,ss2005}. Post recombination, the radiative viscosity of the fluid also reduces, leading to the enhancement of turbulent motion~\cite{jedamzik,jed1998,subb1998}. For scales, where the Reynolds number becomes large, decaying turbulence becomes important and results in the dissipation of magnetic fields. Another process which becomes important post recombination is ambipolar diffusion~\cite{cowling,shu1992}. This process results from the Lorentz force exerted by the magnetic field on the charged particles, thereby accelerating it relative to the neutral particles. The collision between the charged and the neutral particles results in the dissipation of magnetic field and pumps energy into the IGM. The dissipation of magnetic fields caused by this process becomes dominant at lower redshifts. The combined effect of both these processes results in the heating up of the IGM and modifies the thermal and the ionization history of the universe. The impact of non-helical magnetic fields on the temperature and ionization evolution has been studied and constraints on it have been obtained through its impact on CMB temperature and polarization spectra as well as from the EDGES signal~\cite{ss2005,kk2013,chluba2015,kunze2015,min2018,chluba2019}.   \par
In this paper, we have studied the decay of helical magnetic fields in the post-recombination universe via decaying turbulence and ambipolar diffusion and studied its imprints on the CMB. 
The paper is organized as follows: In Section \ref{s1}, we introduce the two point correlation function of the helical magnetic field and set the notation used in this study. In Section~\ref{s2}, we discuss the two dissipation mechanisms, namely, ambipolar diffusion and decaying turbulence. In Section~\ref{s3}, we discuss the added contribution of helicity on the temperature and ionization fraction evolution with further effects on the CMB. Finally, in Section~\ref{s4}, we summarize our main results.

\section{Power spectrum of helical magnetic field}
\label{s1}

Primordial magnetogenesis processes in the early universe generate Gaussian random magnetic fields. These fields evolve adiabaticaly in the linear regime, where the velocity induced by the tangled fields is much less than the Alfv\'en velocity. The two point correlation of the spatial part of the magnetic field, $B(x)$, in the Fourier space is given by~\cite{pog2001,cap2004,tina2005,kk2010},
\begin{equation}
    \langle B_{i}(\vec{k}) B^{\ast}_{j}(\vec{q}) \rangle = (2\pi)^3\delta_{D}^{3}(\vec{k}-\vec{q})\left(P_{S}(k)(\delta_{ij}-\hat{k}_{i}\hat{k}_{j}) - P_{A}(k)i \epsilon_{ijm}\hat{k}_{m} \right)
\end{equation}
where, $P_{S}(k,k_{L},k_{m}) = A_{B}(k/k_{l})^{n_{B}}~W(k,k_{m})$ denotes the power spectrum of the symmetric part with $k_{l}$, $k_{m}$, $n_{B}$ and $W(k,k_{m})$ denoting the pivot scale, diffusion scale, symmetric spectral index, and the Window function, respectively. Similarly, $P_{A}(k,k_{l},k_{m}) = A_{H}(k/k_{l})^{n_{A}} ~W(k,k_{m})$ denotes the power spectrum of the antisymmetric part related to the helicity of the magnetic field with $n_{A}$ denoting the antisymmetric spectral index. The normalized window function, $W(k,k_{m})$, is assumed to be a Gaussian of the form,
\begin{equation}
W(k,k_{m}) = \pi^{-3/2}k^{-3}_{m}\text{e}^{-\left(\frac{k}{k_{m}}\right)^{2}}
\end{equation}
The constants $A_{B}$ and $A_{H}$ are given by,
\begin{align}
    A_{B} &= 4\pi^{7/2}\left(\frac{k_{l}}{k_{m}}\right)^{n_{B}} \rho_{B0} \frac{1}{\Gamma(\frac{n_{B}+3}{2})} \\
    A_{H} &= 4\pi^{7/2}\left(\frac{k_{l}}{k_{m}}\right)^{n_{A}} \rho_{B0} \frac{1}{\Gamma(\frac{n_{B}+3}{2})} \left(\frac{q}{k_{m}}\right)^{n_{B}-n_{A}}
\end{align}
where $\rho_{B0} = \langle B^{2}_{0}(\vec{x})/2 \rangle$ denotes the energy density today smoothed over the diffusion scale. In addition, to satisfy the realizability condition~\cite{kk2010,kk2012}, we have the additional factor of $(q/k_{m})^{n_{B}-n_{A}}$ in the expression for $A_{H}$ where $q = k_{\text{max}}(k_{\text{min}})$ if $n_{A}-n_{B}>0~(n_{A}-n_{B}<0)$, where $k_{\text{max}}$ and $k_{\text{min}}$ denote the maximum and minimum wave number, respectively. For maximally helical magnetic fields, $n_{A}=n_{B}$, for which the additional factor in $A_{H}$ goes away. The dissipation scale, $k_{m}$~\cite{jed1998,subb1998}, is given by,
\begin{equation}\label{km1}
    k_{m}  = 248.60~\left(\frac{B_{0}}{\text{1nG}}\right)^{-1} \text{Mpc}^{-1}
\end{equation}
where, the expression for $k_{m}$ has been computed using the best-fit parameters of Planck 2018~\cite{pl2018}, $\Omega_{m}=0.315 \pm 0.007$, $\Omega_{b}h^2=0.0224 \pm 0.001$, and $H_{0}=67.4 \pm 0.5~\text{km}\text{s}^{-1}\text{Mpc}^{-1}$.
Apart from the diffusion scale, we need to consider the length scale at which the linear regime approximation breaks down. This scale is approximately equivalent to the magnetic Jean's scale, $k_{J}$. For scales, $k > k_{J}$, the magnetic pressure gradients are able to successfully resist gravity to prevent gravitational collapse. Hence, for such scales, non-linear processing prevents the fields from redshifting in a merely expansion driven adiabatic way. Turbulent decay of magnetic field becomes important for scales $ k_{J}< k < k_{m}$. However, in the linear regime i.e. for scales $k < k_{J}$, ambipolar diffusion of magnetic fields dominates. We will discuss both these processes in the next section. 

\section{Dissipation mechanisms of helical magnetic field}
\label{s2}
Post recombination, the number density of ionized particles falls drastically reaching $10^{-4}$ for $z\lesssim 100$~\cite{peeb93}. The temperature of the IGM follows the evolution of $T_{\gamma}$ up to a redshift $\approx 100$ after which it starts falling faster, $\propto a^{-2}$. The dissipation of magnetic field contributes to the heating of the IGM and its ionization, thereby affecting the evolution of the gas temperature and the ionization fraction. The modified temperature evolution is given by~\cite{ss2005,chluba2015,chluba2019}, 
\begin{align}
    \dot{T}_{e} &= -2\frac{\dot{a}}{a}T_{e} + \frac{8 \rho_{\gamma} \sigma_{t}N_{e}}{3 m_{e} N_{tot} c} (T_{\gamma} - T_{e}) + \frac{\Gamma}{1.5k_{B}N_{tot}}
\end{align}
where $N_{tot} = N_{H}(1+f_{He}+X_{e})$ denotes the total number density of particles that are coupled by Coulomb interactions, $T_{\gamma}$ denotes the CMB temperature, $T_{e}$ denotes the gas temperature, and $\sigma_{t}$ denotes the Thomson cross-section. The quantity, $f_{He} \approx Y_{p}/4(1-Y_{p}) \approx 0.079$ for helium mass fraction $Y_{p} = 0.24$; $X_{e}= N_{e}/N_{H}$ denotes the free electron fraction where $N_{e}$ denotes the number density of electrons and $N_{H}$ denotes hydrogen number density. The quantity, $\rho_{\gamma}$ denotes the photon energy density and is given by, $0.26(1+z)^{4}$eV~\cite{chluba2015}. In the above equation, the first term on the R.H.S. represents the decay of the temperature of non-relativistic particles due to expansion. If $T_e<T_{\gamma}$, the second term tends to increase $T_e$. Similarly if $T_e>T_{\gamma}$, this term tends to decrease $T_e$. Thus the second term tends to bring the electron temperature at par with the CMB temperature.
The quantity $\Gamma$ in the third term represents the additional heating due to magnetic fields, which can be either due to a) ambipolar diffusion, $\Gamma_{\text{ambi}}$, b) decaying turbulence, $\Gamma_{\text{turb}}$, or c) a combination of both, $\Gamma_{\text{ambi+turb}}$. We discuss these processes in the following two sub sections.

\subsection{Ambipolar diffusion}

The residual ionization after recombination facilitates ambipolar diffusion wherein the Lorentz force exerted by magnetic fields on the charged particles leads to a velocity difference between the charged and the neutral particles~\cite{cowling,shu1992}. The collisions between the ionized and the neutral particles lead to the dissipation of magnetic fields and in turn heating up of the IGM. The dissipation due to ambipolar diffusion is given by~\cite{cowling,shu1992},
\begin{equation}
    \Gamma_{\text{ambi}} = \frac{(1-X_{e})}{X_{e}  \gamma \rho_{b}^{2}} \frac{\langle{|(\nabla \times B(t,x)) \times B(t,x)|}^{2}\rangle}{16 \pi^{2}}
\end{equation}
where, the mean square Lorentz force, $\langle L^{2}\rangle$ is denoted by $\langle{|(\nabla\times B(t,x))\times B(t,x)|}^{2}\rangle/16\pi^{2}$, $X_{e} = N_{e}/N_{H}$ and $\rho_{b} = m_{H} N_{b}$, where $N_{b}$ denotes the baryon number density and $m_{H}$ denotes the mass of the Hydrogen atom. The quantity, $\gamma$, represents the coupling coefficient and is given by $\langle\sigma \nu \rangle_{HH^{+}}/2 m_{H}$ where, $\langle \sigma \nu \rangle \approx 6.49 \times 10^{-10} (T/K)^{0.375} \text{cm}^{3} \text{s}^{-1}$.  
In Kunze (2011)~\cite{kk2010}, the expression for $\langle L^{2} \rangle$ was estimated to be, 
\begin{align} \label{NHL}
   \langle L^{2} \rangle &= \frac{k^{2}_{m}\rho^{2}_{B0}}{[\Gamma(\frac{n_{B}+3}{2})]^{2}}(1+z)^{10} \int^{\infty}_{0} dw w^{2n_{B}+7}e^{-w^{2}}\int^{\infty}_{0} dv v^{2n_{B}+2}e^{-2v^{2}w^{2}} \int^{1}_{-1} dx e^{2w^{2}vx} \\ &\times \nonumber (1-2vx+v^{2})^{\frac{n_{B}-2}{2}}[1+2v^2+(1-4v^2)x^2-4vx^3+4v^2x^4]
\end{align}
where, $w=k/k_{m}$, $v=q/k$, and $x=\vec{k}.\vec{q}/kq$. This integral is then numerically determined and the final expression is then approximated as, 
\begin{align}
    \langle L^{2} \rangle &\approx 16 \pi^{2} k^{2}_{m} \rho^{2}_{B0} (1+z)^{10} f_{L,NH}(n_{B}+3)
\end{align}
where $f_{L,NH}(n_{B}+3) = 0.8313(1-1.020 \times 10^{-2}(n_{B}+3)) (n_{B}+3)^{1.105}$. We then evaluate $\langle L^{2} \rangle$ for helical magnetic fields using the analysis detailed in Kunze~(2012)~\cite{kk2012},
\begin{align} \label{HL}
    \nonumber
    \langle L^{2} \rangle &=  \frac{k^{2}_{m}\rho^{2}_{B0}}{[\Gamma(\frac{n_{B}+3}{2})]^{2}}(1+z)^{10} \int^{\infty}_{0} dw w^{2n_{B}+7}e^{-w^{2}}\int^{\infty}_{0} dv v^{2n_{B}+2}e^{-2v^{2}w^{2}} \\ &\times \nonumber \int^{1}_{-1} dx e^{2w^{2}vx}  (1-2vx+v^{2})^{\frac{n_{B}-2}{2}}[1+2v^2+(1-4v^2)x^2-4vx^3+4v^2x^4] - \\ \nonumber &\frac{2\rho^{2}_{B0}k^2_{m}}{[\Gamma(\frac{n_{B}+2}{2})]^{2}}\left(\frac{q}{k_{m}}\right)^{2(n_{B}-n_{A})} (1+z)^{10} \int^{\infty}_{0} dw w^{2n_{A}+7}e^{-w^{2}}\int^{\infty}_{0} dv v^{2n_{A}+2}e^{-2v^{2}w^{2}} \\ \nonumber &[ \frac{1}{9} \int^{1}_{-1} dx e^{2w^{2}vx}  (1-2vx+v^{2})^{\frac{n_{A}-1}{2}} (x-v) +\frac{2}{9} \int^{1}_{-1} dx e^{2w^{2}vx}  (1-2vx+v^{2})^{\frac{n_{A}-1}{2}} \\ \nonumber
    &(v+2x-3vx^2) +\frac{2}{9} \int^{1}_{-1} dx e^{2w^{2}vx}  (1-2vx+v^{2})^{\frac{n_{A}-1}{2}} \\ 
    &(4v+2x-6vx^2)]
\end{align}
Using the condition for maximally helical magnetic fields, $n_{A}=n_{B}$, we then numerically evaluate the integral along the lines similar to the non-helical magnetic field case above. The final expression for $\langle L^{2}\rangle$ for helical magnetic fields is given by,
\begin{equation}
    \langle L^{2} \rangle \approx 16 \pi^{2} k^{2}_{m} \rho^{2}_{B0} (1+z)^{10} f_{L,H}(n_{B}+3)
\end{equation}
where $f_{L,H}(n_{B}+3)=0.45(1-0.017(n_{B}+3))(n_{B}+3)^{1.1}$. We use this expression for $f_{L,H}(n_{B}+3)$ to estimate the modified ambipolar diffusion decay of helical magnetic fields.

\subsection{Decaying turbulence}

After recombination, due to the drop in radiative viscosity on scales, $k > k_{J}$, turbulence starts dominating~\cite{jedamzik,jed1998,subb1998}. This leads to non-linear interactions, which through mode coupling leads to the dissipation of magnetic fields from larger to smaller scales until the scale reaches $k_{m}$, after which the dissipation proceeds through viscosity. \par
In previous studies, turbulent decay of non-helical fields has been included in the evolution of temperature of the IGM. Non-helical magnetic fields in the regime of turbulent decay evolve as~\cite{kandu2016,jedamzik},
\begin{equation}
    \rho_{B} = \frac{\rho_{Bi}}{(1+\frac{\tilde{t}}{\tilde{t}_{d}})^{m}} \label{rhoB}
\end{equation}
where, $\rho_{B}$ and $\rho_{Bi}$ denote the flat space magnetic field energy density and the initial value of the flat space magnetic energy density, respectively, and $\tilde{t}$ and $\tilde{t}_{d}$ denote the time in flat space and the relevant dynamical time for decay, respectively. In the above expression, $m = 2(n_{B}+3)/(n_{B}+5)$. However, due to helicity conservation, the value of $m$ for maximally helical magnetic fields takes the value $m=2/3$. Hence, Eq.~\ref{rhoB} is modified to~\cite{kandu2016,jedamzik},
\begin{equation}
    \rho_{B} = \frac{\rho_{Bi}}{(1+\frac{\tilde{t}}{\tilde{t}_{d}})^{2/3}}
\end{equation}
where, the evolution of the magnetic field energy density is independent of the scalar index, $n_{B}$. For non-helical magnetic fields, $\Gamma_{turb}$, is given by the following expression,
\begin{equation}
    \Gamma_{\text{turb,NH}} = \frac{B^{2}_{0}}{8\pi} \frac{3m}{2} \frac{[\text{ln}(1+t_{d}/t_{i})]^{m}}{[\text{ln}(1+t_{d}/t_{i})+\text{ln}(t/t_{i})]^{1+m}} H(t) (1+z)^{4}
\end{equation}
where, $m$ has been defined again as $2(n_{B}+3)/(n_{B}+5)$, $B_{0}$ is the present day magnetic field strength, $H(t)$ is the Hubble parameter, $t_{i}$ is the initial epoch when the decay starts i.e. the epoch of decoupling, and $t_{d}$ denotes the physical decay time scale for the turbulence. The quantity, $t_{d}/t_{i}$, is given by $(k_J/k_{m})^{{n_{B}+5}/2}\approx 14.5 (B_{0}/1\text{nG})^{-1}(k_{m}/1\text{Mpc}^{-1})^{-1}$, where the magnetic Jeans wavenumber, $k_{J}$, is given by,
\begin{equation}
    k_{J}= 14.5^{2/n_{B}+5} \left(\frac{B_{0}}{1\text{nG}}\right)^{-2/(n_{B}+5)}
    \left(\frac{k_{m}}{1\text{Mpc}}\right)^{({n_{B}+3})/({n_{B}+5})}\text{Mpc}^{-1}.
\end{equation}
Due to the change in the evolution equation for the magnetic field energy density, the expression for $\Gamma_{turb}$ for maximally helical magnetic fields is modified to,
\begin{equation}
    \Gamma_{\text{turb,H}} = \frac{B^{2}_{0}}{8\pi}\frac{[\text{ln}(1+t_{d}/t_{i})]^{2/3}}{[\text{ln}(1+t_{d}/t_{i})+\text{ln}(t/t_{i})]^{1+2/3}} H(t) (1+z)^{4}
\end{equation}
Compared to the non-helical case, the exponent $m$ changes to $2/3$ in the helical magnetic field case. This arises from the rate of decay of magnetic field energy density when helicity is conserved. The analytically derived decay rate is also supported by recent numerical simulations conducted by Banerjee and Jedamzik~(2004)~\cite{jedamzik}, Kahniashvili et~al. (2013)~\cite{kahn2013}, and Brandenburg et al. (2015)~\cite{brand2015}.

\begin{figure}
\centering
\includegraphics[scale=0.4]{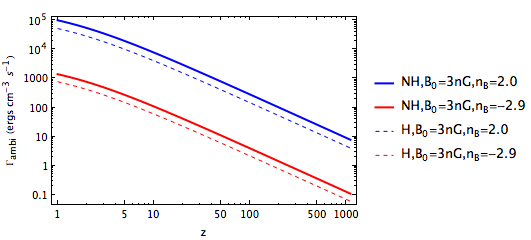}
\includegraphics[scale=0.4]{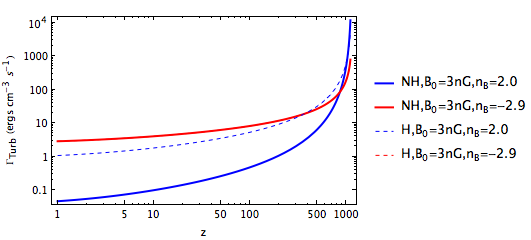}
\caption{The variation of the ambipolar diffusion rate (upper panel) and decaying turbulence rate (lower panel) with redshift, $z$. The magnetic field considered is $B_{0}=3$nG. The solid and dashed lines represent non-helical and helical magnetic fields, respectively. The blue and red lines represent $n_{B}=2.0$ and $n_{B}=-2.9$, respectively.}
\label{fig0}
\end{figure}

Next, we make use of these modified expressions for ambipolar diffusion and turbulent decay to evaluate the temperature and ionization fraction evolution. To this end, we use the extension of the  \emph{RECFAST++} code~\cite{chluba2010}, which includes the contribution from magnetic field. We have modified the same to include the effect of helicity.  We have also plotted the variation of the ambipolar diffusion and decaying turbulence rate with redshift, $z$ for both non-helical and helical magnetic fields in Fig.~\ref{fig0}. Further, we also study the implication of helicity on the CMB temperature and polarization anisotropies using \emph{CAMB}~\cite{lewis2000} in the following section. 

\section{Results and Discussion} \label{s3}

\subsection{Evolution of temperature and ionization fraction}

\begin{figure}[htbp]
    \centering
    \includegraphics[scale=0.3]{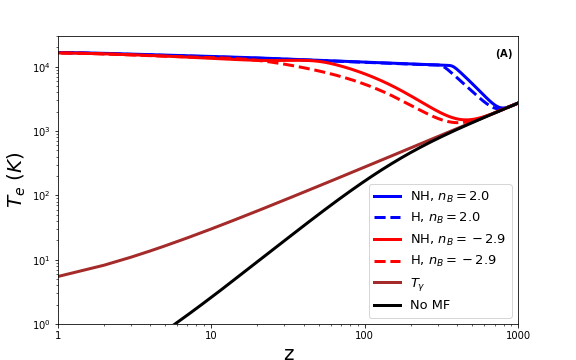}
    \includegraphics[scale=0.3]{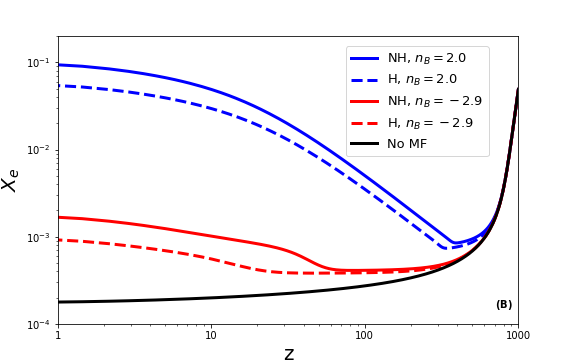}
    \includegraphics[scale=0.3]{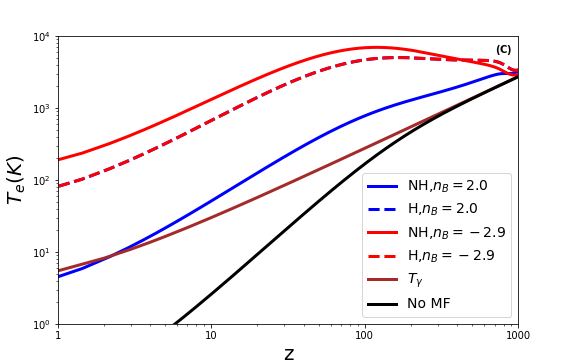}
    \includegraphics[scale=0.3]{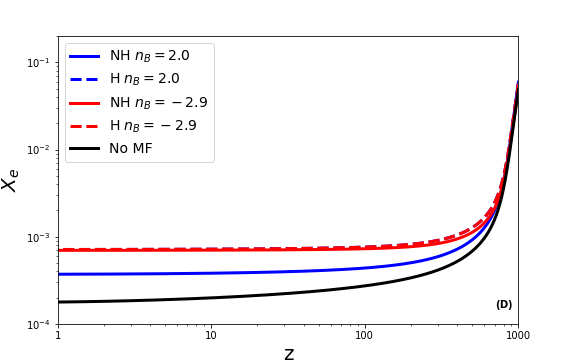}
    \includegraphics[scale=0.3]{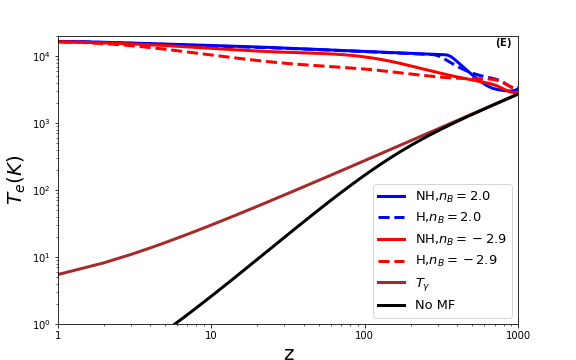}
    \includegraphics[scale=0.3]{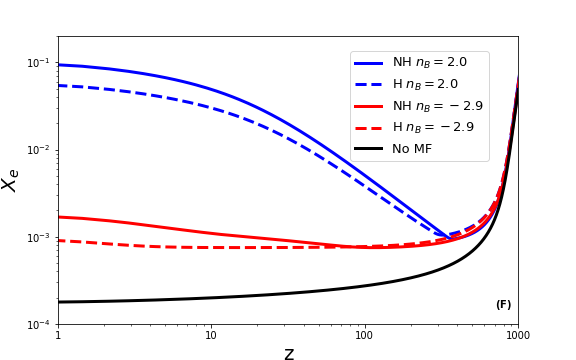}
    \caption{\emph{Left panel:} Temperature evolution with redshift due to the additional effect of ambipolar diffusion (A), decaying turbulence (C) and the combined effect of both (E) for non-helical as well as helical magnetic fields. The blue solid and dashed line represent the contribution due to non-helical and helical magnetic fields, respectively for $n_{B}=2.0$. Similarly, the red solid and dashed lines represent the contribution due to non-helical and helical magnetic fields, respectively for $n_{B}=-2.9$. The magnetic field considered is $B_{0}=3$nG. The black solid line represents the CMB temperature and the brown solid line represents the gas temperature without magnetic field. \emph{Right panel:} Ionization fraction evolution with redshift with similar details as given in the left panel.}
    \label{fig1}
\end{figure}

In  Fig.~\ref{fig1}, we have plotted the temperature evolution (left panel) and the evolution of ionization fraction (right panel) due to ambipolar diffusion (upper panel), turbulent decay (middle panel) and the combined effect of both for non-helical and maximally helical magnetic fields~(lower panel). We can infer from the figure that under certain conditions the contribution of the antisymmetric part of the magnetic field power spectrum can lead to a perceivable deviation from the non-helical case in both the baryon temperature evolution as well as the ionization fraction evolution. 

As seen from upper left panel of Fig.~\ref{fig1}, we note that there is a decrease in the temperature of the gas in the case of helical field as compared to the non-helical case for certain intermediate values of redshift, $z >12$, for both the blue spectrum~($n_{B}=2.0$) as well as the nearly scale invariant spectrum~($n_{B}=-2.9$). In the case of the nearly scale invariant spectrum ($n_{B}=-2.9$), the deviation of the gas temperature (between the helical and the non-helical case at intermediate redshifts) is more than that compared to the $n_B=2$ case. These curves however, merge with each other at lower redshifts showing that although the magnetic field results in a change in gas temperature at low redshifts (as compared to the case when magnetic field is not considered), it is independent of whether we consider helical fields or non helical fields and whether we consider a blue spectrum or a nearly scale invariant one. \par

The ionization fraction evolution with redshift (upper right panel) also shows a similar trend with the presence of magnetic field leading to an increase in the degree of ionization. The helical contribution however results in an ionization fraction which is less than that in the case of non-helical fields. This feature is exhibited for both the blue spectrum as well as the nearly scale invariant spectrum.  Both the behaviour of temperature and ionization fraction evolution can be understood from the change in the expression of $\langle L^2 \rangle$ (See Eq.~\ref{NHL} and Eq.~\ref{HL}), which has an additional negative contribution from the antisymmetric part in the case of helical magnetic fields. 

Another important aspect which is worth noting is that while the deviation in the temperature vanishes at low redshifts for the two cases (of helical and non-helical magnetic fields), the difference in the ionization fraction for the case of helical and non-helical field is clearly perceivable even at low redshifts for both the blue spectrum as well as the nearly scale invariant spectrum. As we shall see this leads to a change in the polarization spectrum of the CMB. The above behaviour has been estimated for $B_{0}=3$nG and for higher magnetic fields, the change will be higher.

Next, equating $\Gamma = \Gamma_{turb}$, we can infer that the contribution of the antisymmetric part is independent of the spectral index and leads to higher values of temperature and ionization fraction as a function of redshift compared to the non-helical magnetic field for higher values of $n_{B}$. This can also be inferred from the middle panel of Fig.~\ref{fig1} where the temperature evolution for the helical magnetic field is shifted to higher values with respect to the temperature evolution for the non-helical magnetic field when $n_{B}=2.0$. However, as $n_{B}$ reduces to $-2.9$, the non-helical magnetic field induced temperature exceeds the helical magnetic field induced temperature. This can be understood by comparing the variation of $\Gamma_{turb,NH}$ and $\Gamma_{turb,H}$ for different values of $n_{B}$. Similarly, the ionization fraction evolution shows that the helical magnetic field induced ionization fraction as a function of redshift is larger than that induced due to non-helical magnetic fields for $n_{B}=2.0$ but for lower spectral indices, the difference diminishes with the evolution almost overlapping for $n_{B}=-2.9$. Next, in the lower panel of Fig.~\ref{fig1}, we show the combined effect of ambipolar diffusion and turbulence on the temperature and the ionization fraction evolution. We can see that the magnetic field decay due to turbulence dominates at high redshifts with ambipolar diffusion taking over at lower redshifts. With the temperature and the ionization fraction evolution in place, we now proceed to study the effect of the helical magnetic field dissipation on the CMB. 

\subsection{Effect on CMB} 

\begin{figure}[htbp]
    \centering
    \includegraphics[scale=0.3]{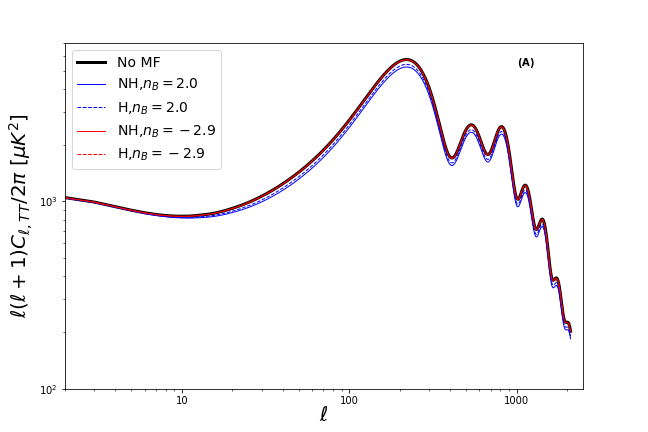}
    \includegraphics[scale=0.3]{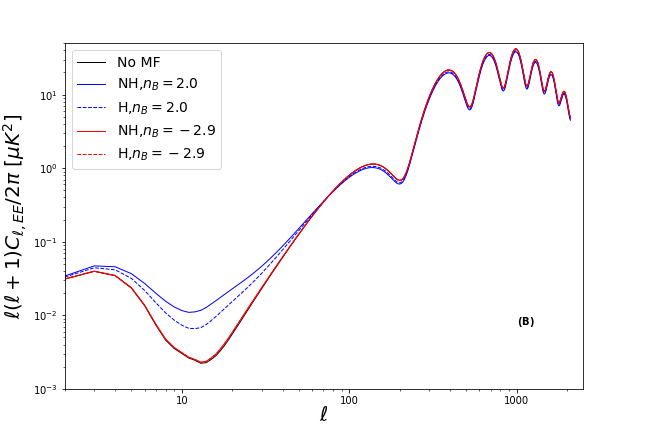}
    \includegraphics[scale=0.3]{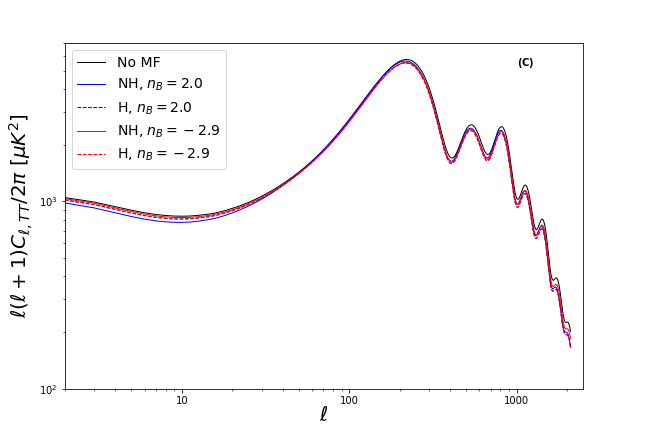}
    \includegraphics[scale=0.3]{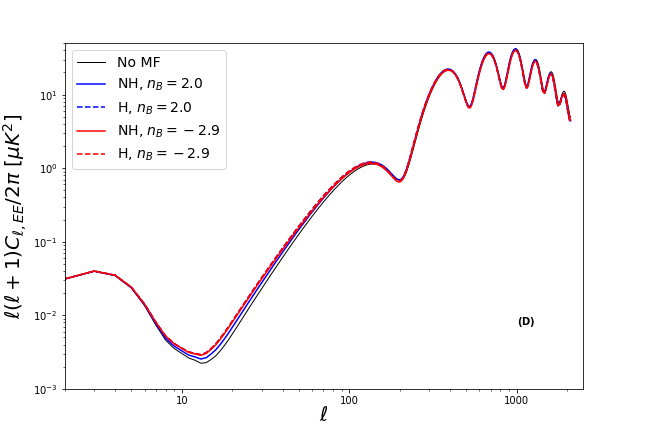}
    \caption{\small{(A) and (C)  denote the CMB temperature anisotropy spectrum where the x-axis denotes the multipole, $\ell$ with the effect of ambipolar diffusion and decaying turbulence, respectively. Similarly,  (B), and (D) denote the CMB polarization anisotropy spectrum where the x-axis denotes the multipole, $\ell$ with the effect of ambipolar diffusion and decaying turbulence, respectively. The magnetic field considered here is $B_{0}=3$nG. The blue solid and dashed lines represent contribution of non-helical magnetic fields and maximally helical magnetic fields, respectively, for $n_{B}=2.0$. Keeping everything the same, red solid and dashed lines represent $n_{B}=-2.9$. The black solid line represents the temperature anisotropy spectrum with no magnetic field. }}
\label{fig2}
\end{figure}

The effect on CMB through ambipolar diffusion and turbulent decay of non-helical magnetic fields has been studied before and constraints on the magnetic field have also been derived~\cite{kunze2015,chluba2015,chluba2019}. In this paper, we study the  effect of post-recombination dissipation of helical magnetic fields on the CMB. Fig.~\ref{fig2} shows the effect of $3$nG magnetic field (both non-helical as well as maximally helical) on the temperature and polarization anisotropy of the CMB. As is clear from the figure, while the effect in CMB temperature is small, the effect on CMB polarization anisotropy is clearly seen at least in a certain range of $\ell$-values.

\begin{figure}[htbp]
    \centering
   \includegraphics[width = 0.5\textwidth]{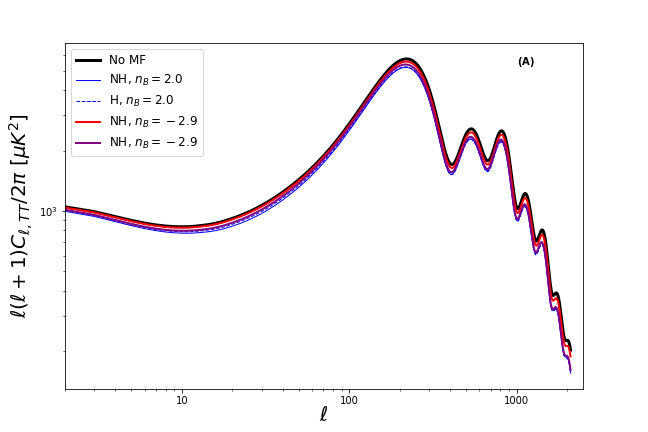}
   \includegraphics[width = 0.5\textwidth]{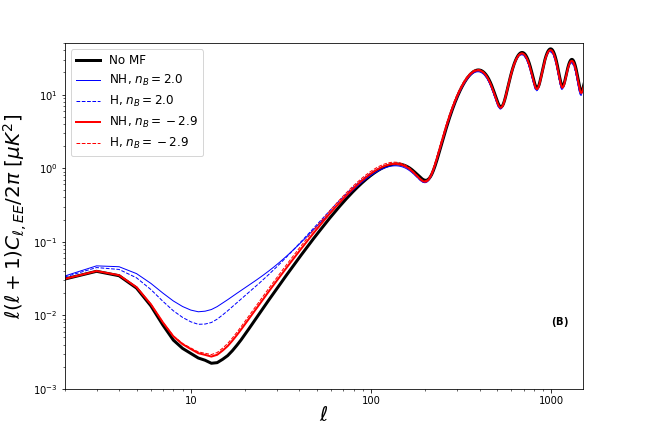}
   \includegraphics[width = 0.5\textwidth]{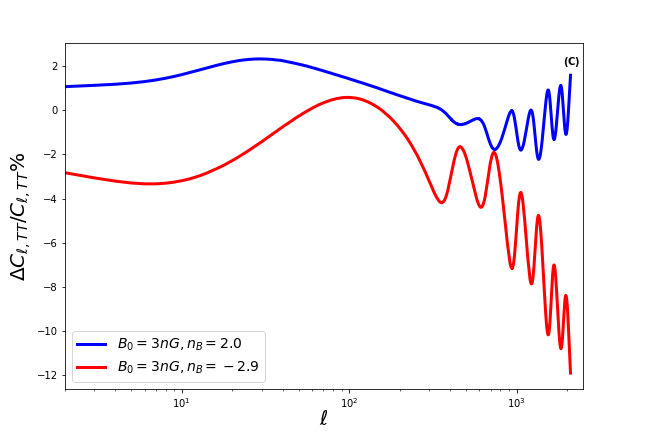}
   \includegraphics[width = 0.5\textwidth]{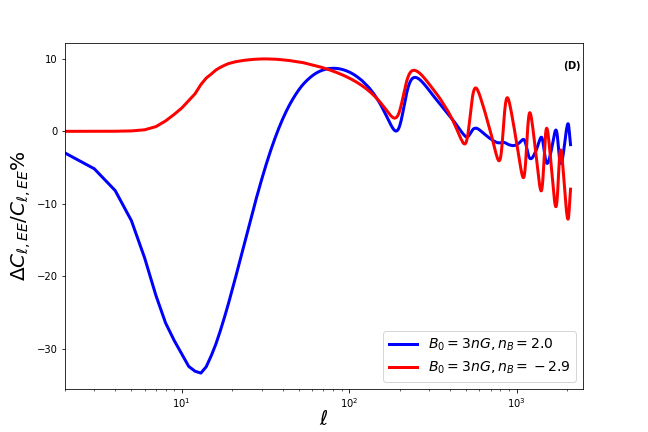}
    \caption{\small{(A) and (B) denote the CMB temperature and polarization anisotropy spectrum, respectively where the x-axis denotes the multipole, $\ell$ with the combined effect of ambipolar diffusion and decaying turbulence. (C) and (D) denote the  percentage difference in $C_{\ell,TT}$ and $C_{\ell,EE}$, respectively,  between helical and non-helical magnetic field with the x-axis denoting $\ell$. $\Delta C_{\ell}/C_{\ell}$ denotes $C_{\ell,H}-C_{\ell,NH}/C_{\ell,NH}\times100$, where $C_{\ell,H}$ and $C_{\ell,NH}$ represent the CMB temperature anisotropy due to helical and non-helical magnetic field, respectively. It includes the combined effect of ambipolar diffusion and decaying turbulence in the calculation of the percentage difference in CMB temperature anisotropies. The magnetic field considered here is $B_{0}=3$nG. The blue solid and dashed lines represent contribution of non-helical magnetic fields and maximally helical magnetic fields, respectively, for $n_{B}=2.0$. Keeping everything the same, red solid and dashed lines represent $n_{B}=-2.9$. The black solid line represents the temperature anisotropy spectrum with no magnetic field. }}
    \label{fig3}
\end{figure}

In Fig.~\ref{fig2}A and Fig.~\ref{fig2}C, we have shown how the CMB temperature is affected by the dissipation of magnetic filed (maximally helical denoted by dashed line and non-helical magnetic fields denoted by solid lines) through ambipolar diffusion and turbulent decay, respectively. Fig.~\ref{fig3} incorporates the combined effect of ambipolar diffusion and turbulent decay. In the case of ambipolar diffusion, the effect of helicity shifts the temperature anisotropy spectrum to slightly higher values compared to the case of non-helical magnetic fields. The maximum value of the shift is $\approx 3\%$ for $n_{B}=2$ and decreases further with decreasing spectral index. The variation is negligible for the spectral index, $n_{B}=-2.9$. Further, in the case of decaying turbulence, the helical magnetic field shifts the temperature anisotropies to slightly lower values compared to non-helical magnetic fields at higher $\ell$s with the shift being more for $n_{B}=-2.9$ than $n_{B}=2.0$. This behaviour is different at lower $\ell$s for $n_{B}=2.0$ compared to $n_{B}$ = -2.9. In addition, as decaying turbulence for helical magnetic fields is independent of the spectral index, we see that the temperature anisotropy spectra for all values of the spectral index coincide. In order to resolve the difference between the helical and non-helical magnetic fields, in Fig.~\ref{fig3}C, we have plotted the percentage difference in the CMB temperature anisotropy due to helical and non-helical magnetic field including both the ambipolar diffusion effect and the decaying turbulence effect. We can infer from the figure that for $n_{B}= -2.9$, the effect of decaying turbulence dictates the behaviour as the effect of ambipolar diffusion in this case is negligible. On the other hand, for $n_{B}=2.0$, the combination results in the helical magnetic field leading to higher temperature anisotropies than the non-helical magnetic field for most values of $\ell$ as the effect of ambipolar diffusion dominates for this case. \par

The difference in polarization anisotropy of the CMB between the helical vs non-helical cases is more prominent than the temperature anisotropy. On the right panel of Fig.~\ref{fig2}, we have shown the polarization anisotropy in CMB due to ambipolar diffusion (Fig.~\ref{fig2}B), turbulent decay (Fig.~\ref{fig2}D) and in Fig.~\ref{fig3}B,  we have shown the combination of both the effects.  As in the temperature anisotropy case, we have also shown the percentage difference between the helical and the non-helical magnetic field in Fig.~\ref{fig3}D. In the case of ambipolar diffusion, the helical field shifts the polarization anisotropy spectrum to lower values and the shift decreases with decreasing $n_{B}$. This can be understood from the behaviour of the temperature and ionization fraction evolution, which shows lower values for helical magnetic fields compared to non-helical magnetic fields. The decrease in the ionization fraction leads to a decrease in the polarization anisotropy. We also see that the effect of turbulent decay on polarization is smaller than ambipolar diffusion. In the case of decaying turbulence,  helical magnetic field induced polarization anisotropy has higher values than the non-helical case except at high $\ell$s~(Fig.~\ref{fig2}D).  

The net effect of ambipolar diffusion and turbulent decay due to both non-helical and helical magnetic fields  on CMB polarization is shown in Fig.~\ref{fig3}D. We find that for the nearly scale invariant spectrum ($n_{B}=-2.9$), the polarization anisotropy in CMB is dictated by decaying turbulence as the effect of ambipolar diffusion is negligible in this case. On the other hand, for $n_{B}=2.0$, the combined plot shows that the CMB polarization power spectrum closely resembles that due to the ambipolar diffusion as the effect of decaying turbulence dictates the behaviour for only a small range of $\ell$. In order to highlight how different the effect is between these two cases, the percentage difference in the polarization anisotropy for these two cases are plotted in Fig.~\ref{fig3}D.

We note that for the blue spectrum ($n_B=2$) there is a dip of $~35\%$ in the case of helical magnetic field induced polarization anisotropies compared to the non-helical magnetic field case for small values of $\ell$ roughly between $5$ and $20$. For $40 \le \ell \le 450$ the effect of the helical field is more, reaching a maximum of about $9\%$. For larger $\ell$, the difference is not much and remains within $5\%$. For $n_{B}=-2.9$, the induced polarization anisotropies due to helical field is not significantly different from the non-helical one, the largest difference being less than about $10\%$.

\begin{figure}[htbp]
    \centering
    \includegraphics[width = 0.5\textwidth]{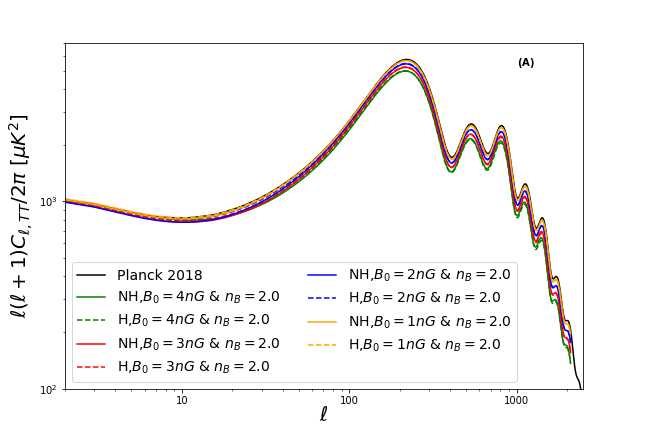}
    \includegraphics[width = 0.5\textwidth]{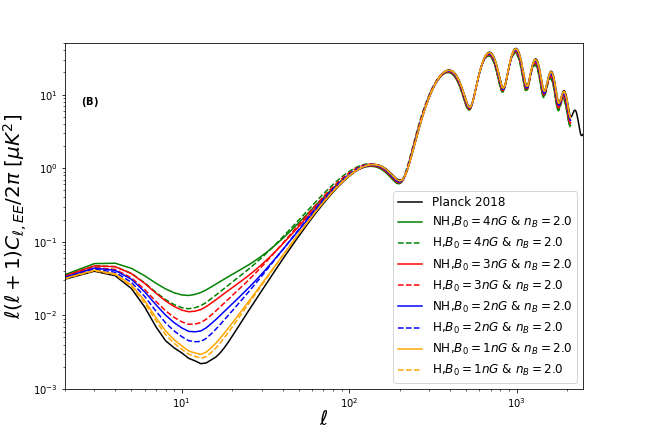}
   \includegraphics[width = 0.5\textwidth]{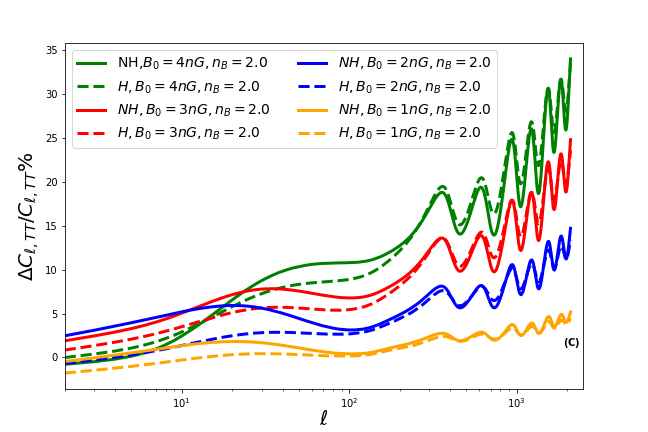}
    \includegraphics[width = 0.5\textwidth]{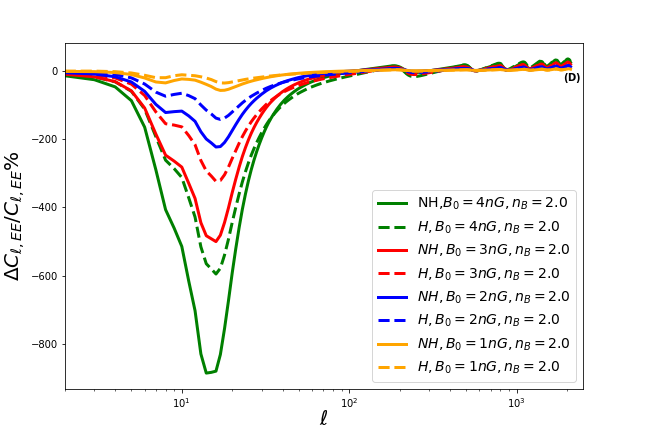}
 \caption{\small{ (A) and (B) denote the CMB temperature and polarization anisotropy spectra for $n_{B}=2.0$ , respectively, where the x-axis denotes the multipole, $\ell$.  (C) and (D) denote the percentage difference in $C_{
\ell,TT}$ and $C_{\ell,EE}$, respectively estimated through \emph{CAMB} with respect to  Planck 2018 temperature and polarization anisotropies with $\Delta C_{\ell}/C_{\ell}$ denoting $(C_{\ell,pl}-C_{\ell,NH or H})/C_{\ell,pl}~\times 100$ where $C_{\ell,pl}$ represents the Planck 2018 CMB anisotropy and $C_{\ell,NH or H}$ denotes that calculated from \emph{CAMB} for non-helical or helical magnetic fields. The Planck 2018 observations are indicated via the black solid line. All the plots include the combined effect of ambipolar diffusion and decaying turbulence with different strengths of magnetic field. The green, red, blue, and orange solid lines represent $B_{0}=4$nG, $B_{0}=3$nG, $B_{0}=2$nG, and $B_{0}=1$nG, respectively. The corresponding dashed lines represent helical magnetic fields.}}  
   
    \label{fig4}
\end{figure}

\begin{figure}[htbp]
    \centering
    \includegraphics[width = 0.5\textwidth]{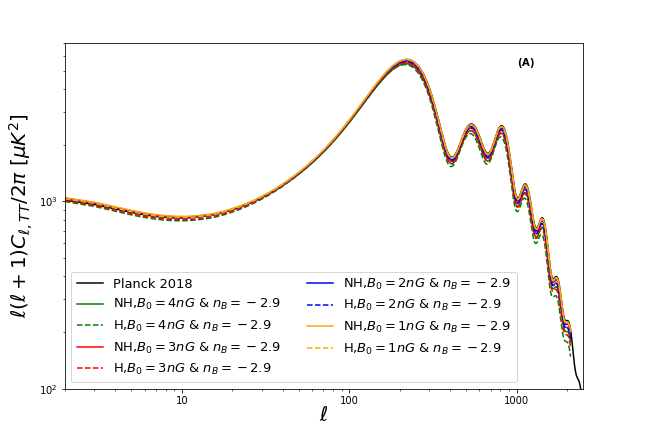}
   \includegraphics[width = 0.5\textwidth]{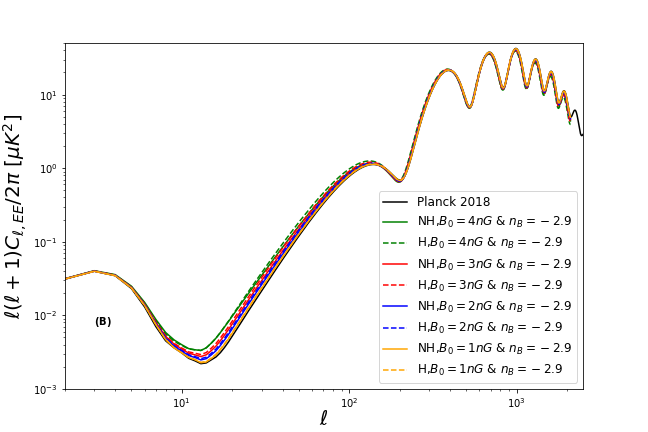}
    \includegraphics[width = 0.5\textwidth]{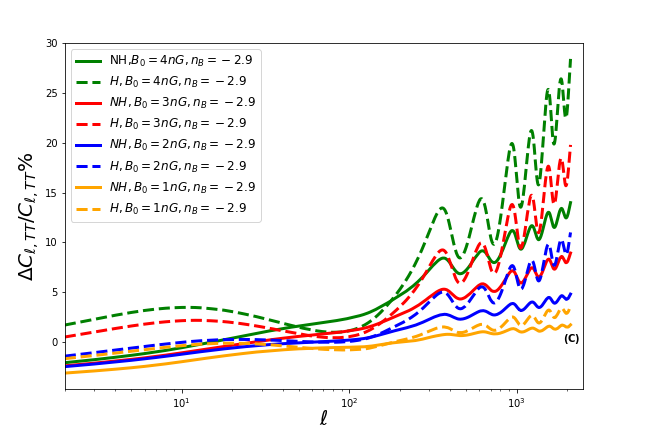}
    \includegraphics[width = 0.5\textwidth]{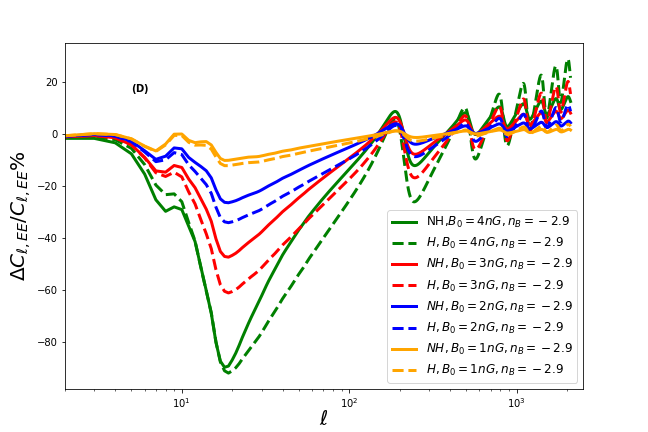}
 \caption{\small{ (A) and (B) denote the CMB temperature and polarization anisotropy spectra for $n_{B}=-2.9$, respectively, where the x-axis denotes the multipole, $\ell$.  (C) and (D) denote the percentage difference in $C_{
\ell,TT}$ and $C_{\ell,EE}$, respectively estimated through \emph{CAMB} with respect to  Planck 2018 temperature and polarization anisotropies with $\Delta C_{\ell}/C_{\ell}$ denoting $(C_{\ell,pl}-C_{\ell,NH or H})/C_{\ell,pl}~\times 100$ where $C_{\ell,pl}$ represents the Planck 2018 CMB anisotropy and $C_{\ell,NH or H}$ denotes that calculated from \emph{CAMB} for non-helical or helical magnetic fields. The Planck 2018 observations are indicated via the black solid line. All the plots include the combined effect of ambipolar diffusion and decaying turbulence with different strengths of magnetic field. The green, red, blue, and orange solid lines represent $B_{0}=4$nG, $B_{0}=3$nG, $B_{0}=2$nG, and $B_{0}=1$nG, respectively. The corresponding dashed lines represent helical magnetic fields.}}  
  \label{fig5}
  \end{figure}

In addition, we have shown in Fig.~\ref{fig4} and Fig.~\ref{fig5}, the CMB temperature and anisotropy spectrum obtained from Planck 2018~\cite{pl2018} to compare with the CMB temperature and polarization anisotropies computed using \emph{CAMB} for different magnetic field strengths and spectral indices ($n_{B}=2.0$ and $n_{B}=-2.9$). In Fig~\ref{fig4}C, D and Fig~\ref{fig5}C, D  
, we have plotted the absolute percentage deviation of our models from that of Planck 2018. We have included the combined effect of ambipolar diffusion and turbulent decay to compute these plots. It can be inferred from Fig.~\ref{fig4} that the maximally helical magnetic field generated temperature anisotropy spectrum for $n_{B}=2.0$ is closer to the Planck 2018 spectrum as compared to that due to the non-helical magnetic field at most $\ell$s for the same magnetic field strength. This was also seen in Fig.~\ref{fig2} and Fig.~\ref{fig3} where we showed that ambipolar diffusion dominates the behaviour for $n_{B}=2.0$. This implies that the constraint obtained on the magnetic field is relaxed when helicity is included for $n_{B}=2.0$. On the other hand for $n_{B}=-2.9$, the CMB anisotropy spectra due to helical magnetic field is farther from the Planck 2018 spectra compared to the non-helical case for most values of $\ell$ (See Fig.~\ref{fig5}). Further, as the magnetic field is reduced, the spectra approaches the Planck 2018 spectra with magnetic field between $1-2$nG showing the least deviation for both values of $n_{B}$.  However, for $n_{B}=2.0$, this constraint is relaxed and for $n_{B}=-2.9$, it is tightened as we saw above. We note here that the constraints discussed are qualitative and have been inferred from the percentage difference plots.

Before we end this section, we would like to present a comparison of our results for the non-helical magnetic field case with previous results such as Chluba et al.~(2015)~\cite{chluba2015} and Kunze and Komatsu~(2015)~\cite{kunze2015} and study the variation of our results with an alternative model of damping profile discussed in Paoletti~et~al.~(2019)~\cite{pao2019}. The overall trend of the change in the CMB anisotropies due to the ambipolar diffusion and decaying turbulence effect of non-helical magnetic fields match with that provided by Chluba et al.~(2015)~\cite{chluba2015} where $B_{0}=3$nG and $n_{B}=-2.9$ and $n_{B}=3.0$ had been considered. The amplitudes of the percentage difference in $C_{\ell}$ with and without magnetic field calculated in our analysis are also equivalent with their work for $n_{B}=-2.9$. 
The constraints on the magnetic field obtained in Chluba~et.al.~(2015)~\cite{chluba2015} namely, $B_{0}<1.1$nG for $n_{B}=-2.9$ match with the qualitative constraints inferred from the plots in this work. In addition, the latter is also of similar order to the constraints obtained by Kunze and Komatsu~(2015)~\cite{kunze2015}, wherein they employed Recfast++ with CLASS to study the variation of CMB anisotropies with non-helical magnetic fields.

\begin{figure}[htbp]
\centering
\includegraphics[scale=0.3]{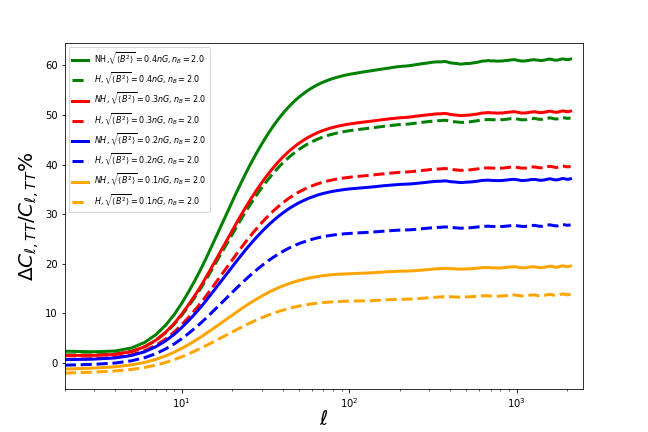}
\includegraphics[scale=0.3]{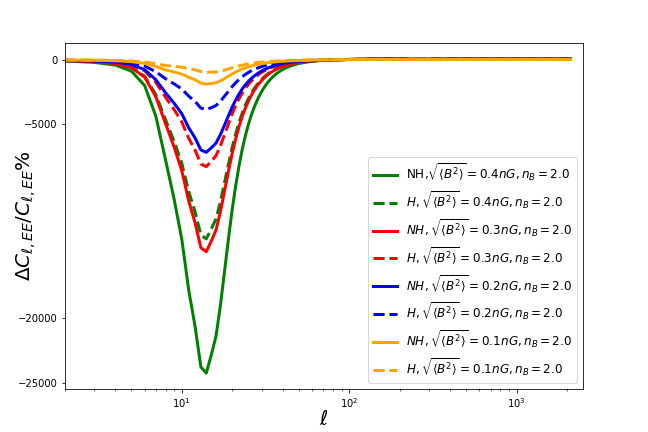}
\includegraphics[scale=0.3]{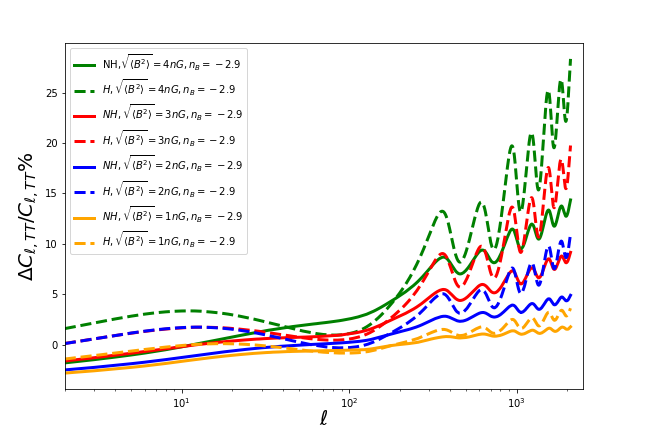}
\includegraphics[scale=0.3]{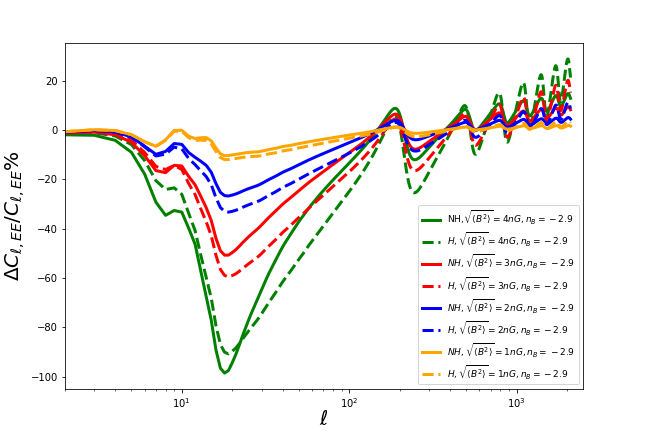}
\caption{The upper and the bottom two panels denote the percentage difference with respect to Planck 2018 results for $n_{B}=2.0$ and $n_{B}=-2.9$ using the new damping scale. The solid and dashed lines represent non-helical and helical magnetic fields, respectively. The magnetic fields considered for the upper two panels are namely, $0.4$nG,$0.3$nG, $0.2$nG, and $0.1$nG. On the other hand, the magnetic fields considered for the lower two panels are namely, $4$nG,$3$nG, $2$nG, and $1$nG.}
\label{fignew}
\end{figure}

We now compare our results for an alternative damping scale as given below with  the results obtained by Paoletti~et~al.~(2019)~\cite{pao2019},
\begin{equation} \label{km}
    k_{m} = \frac{\sqrt{5.5 \times 10^{4}}}{\sqrt{\langle B^{2} \rangle/nG}} \frac{(2\pi)^{(n_{B}+3)}/2}{\sqrt{\Gamma[n_{B}+5)/2]}} 
    \sqrt{h\frac{\Omega_{b}h^{2}}{0.022}} \text{Mpc}^{-1}
\end{equation}
wherein the damping profile is given by,
\begin{equation}
    \langle B^{2} \rangle = \frac{1}{2\pi^{2}} \int_{0}^{k_{m}} dk k^{2} 2P_{B} (k)
\end{equation}
The change in the dissipation scale does not affect the overall trend seen previously as can be inferred from Fig.~\ref{fignew}. In other words, ambipolar diffusion still dominates for $n_{B}=2.0$ and decaying turbulence dominates for $n_{B}=-2.9$. However, the amplitude of change in $C_{\ell,TT}$ and $C_{\ell,EE}$ alters considerably for $n_{B}=2.0$ . A similar change is not observed in the case of $n_{B}=-2.9$.  The temperature and ionization fraction evolution with redshift, z, also registers a change for $n_{B}=2.0$ but not for $n_{B}=-2.9$. The constraints obtained for $\sqrt{ \langle B^{2} \rangle}$ is considerably tightened compared to $B_{0}$. Paoletti et al. (2019)~\cite{pao2019} using MCMC had obtained $\sqrt{ \langle B^{2} \rangle} < 0.06$nG for $n_{B}=1.0$ and $\sqrt{ \langle B^{2} \rangle} < 1.06$nG for $n_{B}=-2.9 $, which matches with our qualitative estimates. 

In addition, including the helicity does not alter the overall trend as seen for the previous damping scale; however, the amplitude of the difference between the non-helical and helical magnetic fields increases considerably for the ambipolar diffusion case as the rate is proportional to $k^{2}_{m}$ and the new damping scale is greater than the old damping scale by $(2\pi)^{(n_{B}+3)/2}/\sqrt{\Gamma[n_{B}+5)/2]}$. Even for magnetic field strength as low as $\sqrt{\langle B^{2} \rangle} = 0.4.$nG, the amplitude of the percentage difference between the $C_{\ell,TT,H}$ and $C_{\ell,TT,NH}$  is $\approx 15~\%$ for $n_{B}=2.0$. As we have stated previously, no difference from the previous result is observed for $n_{B}=-2.9$. We have also plotted the percentage difference in $C_{\ell,TT}$ and $C_{\ell,EE}$ with respect to Planck 2018 results for the new dissipation scale in Fig.~\ref{fignew} where the modification due to the new dissipation scale is apparent for $n_{B}=2.0$.

\section{Conclusions} \label{s4}

In this paper, we have studied the post recombination decay of maximally helical magnetic fields. After recombination, the ionization fraction drops down to an order of $10^{-4}$, which enables the magnetic field to decay via ambipolar diffusion. Similarly, due to the decrease in viscosity at length scales less than the magnetic Jeans length, turbulence dominates and leads to the decay of magnetic field. Both these processes combined leads to the modification of the temperature and ionization fraction evolution of the IGM. \par
In contrast with earlier studies conducted on this subject, we have studied the additional contribution of helicity of the magnetic field. There are various parity breaking magnetogenesis models that predict helical magnetic fields, which have larger coherence length and a slower decay rate compared to non-helical magnetic fields. However, their decay post-recombination and subsequent implications on CMB have not been considered before. After taking into account the helical component of the magnetic field, the ambipolar diffusion term and the turbulent decay term is modified. 
By considering only the dissipation due to ambipolar diffusion, we find that for maximally helical magnetic fields, which have $n_{A}=n_{B}$, the values of temperature and ionization fraction are smaller at certain redshifts compared to the case with non-helical magnetic fields as indicated in the upper panel of Fig.~\ref{fig1}. This also affects the CMB temperature and anisotropy results, which registers a non-negligible change in the temperature and polarization as depicted in the upper panel of Fig.~\ref{fig2}. Similarly, we evaluated the effect of decaying turbulence and the combined effect of ambipolar diffusion and decaying turbulence on the CMB anisotropies as depicted in the lower panel of Fig.~\ref{fig2} and Fig.~\ref{fig3}, respectively. \par

We can infer from Fig.~\ref{fig4} and  Fig.~\ref{fig5} that the presence of helicity relaxes the constraints on the magnetic field for $n_{B}=2.0$ and tightens the constraints for $n_{B}=-2.9$, respectively. This is because ambipolar diffusion dominates the behaviour for the former and turbulent decay dominates the latter.  It implies that ambipolar diffusion dictates the constraints for higher spectral indices compared to turbulent decay which dictates the behaviour for lower spectral indices. The constraints discussed here are qualitative in nature and have not been derived using any sampler such as MCMC compared to the analysis performed by Paoletti~et al~(2019)~\cite{chluba2019}, wherein constraints on non-helical magnetic fields were derived after performing MCMC. However, the overall trend and the qualitative constraints discussed here match with their quantitative constraints.  After including the antisymmetric part of the magnetic field into the computation, we infer from the percentage difference plots that there is a small change in the constraints.  However, evidence suggests the presence of helical magnetic fields in various astrophysical systems; hence, the study of their impact on existing observations is important. Further, this could pave the way for another independent constraint on cosmic magnetic fields.

\section*{Acknowledgements}
SJ and TRS acknowledge the facilities at IUCAA
Centre for Astronomy Research and Development
(ICARD), University of Delhi. RS would like to thank Professor Kandaswamy Subramanian for insightful discussions. The research of SJ is supported by INSPIRE Fellowship (IF160769), DST India. TRS acknowledges the project grant from SERB, Govt. of India (EMR/2016/002286). 

\bibliographystyle{ws-ijmpd}
\bibliography{reference}

\end{document}